\documentclass[12pt]{article}
\usepackage{amssymb,amsmath}

\newcommand{\eqn}[1]{(\ref{#1})}
\def\appendix#1{\addtocounter{section}{1}\setcounter{equation}{0}
\renewcommand{\thesection}{\Alph{section}}
\section*{%Appendix~
\thesection\protect\indent \parbox[t]{11.715cm} {#1}}
\addcontentsline{toc}{section}{Appendix\thesection\ \ \ #1} }
\newcommand{\complex}{{\mathbb C}} %% complex numbers
\newcommand{\zed}{{\mathbb Z}} %% integers
\newcommand{\id}{{\mathbb I}} %%operatore 1

\def\nn{\nonumber}
\newcommand{\tr}[1]{\:{\rm tr}\,#1}

\def\e{{\,\rm e}\,}
\def\ii{{\,\rm i}\,}

\newcommand{\be}{\begin{equation}}
\newcommand{\ee}{\end{equation}}
\newcommand{\beq}{\begin{equation}}
\newcommand{\eeq}{\end{equation}}
\newcommand{\bea}{\begin{eqnarray}}
\newcommand{\eea}{\end{eqnarray}}

\newcommand{\del}{\partial}
\newcommand{\bra}[1]{\left\langle #1\right|}
\newcommand{\ket}[1]{\left| #1\right\rangle}

\newcommand{\delslash}{\partial\!\!\!/}

\begin{document}
\begin{titlepage}
\begin{flushright}

\baselineskip=12pt
DSF--31--2006\\ hep--th/0610023\\
\hfill{ }\\
September 2006
\end{flushright}

\begin{center}

\baselineskip=24pt

{\Large\bf Internal Space for the Noncommutative Geometry Standard
Model and Strings}

\baselineskip=14pt

\vspace{1cm}

{{\bf Fedele~Lizzi$^{1,2}$}}
\\[6mm]
 $^{1}${\it Dipartimento di Scienze Fisiche, Universit\`{a} di
Napoli {\sl Federico II}}\\ $^{2}${\it INFN, Sezione di Napoli}\\
Monte S.~Angelo, Via Cintia, 80126 Napoli, Italy\\ {\small\tt
fedele.lizzi@na.infn.it}

\end{center}

\vskip 2 cm

\begin{abstract}
In this paper I discuss connections between the noncommutative
geometry approach to the standard model on one side, and the
internal space coming from strings on the other. The standard model
in noncommutative geometry is described via the spectral action. I
argue that an internal noncommutative manifold compactified at the
renormalization scale, could give rise to the almost commutative
geometry required by the spectral action. I then speculate how this
could arise from the noncommutative geometry given by the vertex
operators of a string theory.

\end{abstract}

\end{titlepage}

\section{Introduction}

In this paper I present some speculative conjectures linking the
programme started in~\cite{ConnesLott}, where the ordinary
\emph{standard model} of particle physics is seen as particular
noncommutative geometry~\cite{Connesbook,ticos}, with some
features of \emph{string
theory}~\cite{GreenSchwarzWitten,Polchinskibook}. The standard
model in noncommutative geometry is described by the
\emph{spectral action}~\cite{chamseddineconnes1,
chamseddineconnes2} of a noncommutative geometry. All topological
information of a Hausdorff space is contained in the commutative
algebra of continuous complex-valued functions defined over it,
and a noncommutative space is the generalization to noncommutative
$C^*$-algebras. Although the most familiar examples are
deformation of ordinary manifolds (such as the Moyal plane or the
noncommutative torus), noncommutative geometry has a much wider
scope, and there are examples of noncommutative geometries (such
as the space of Penrose tilings~\cite{Connesbook}) which are not
deformations of a ordinary commutative manifolds.

While the topology is encoded in a $C^*$-algebras, much of the
geometry is encoded in the (generalized) Dirac operator, a self
adjoint operator which acts on the Hilbert space on which the
algebra is represented as bounded operators. With the Dirac operator
is possible to define a differential calculus representing also
forms as operators, and it is possible to build the action, which is
the measure of the fluctuations of the geometry. The initial great
success of this programme has been the fact that, by considering as
geometry an \emph{almost commutative} geometry composed of ordinary
spacetime and an internal zero-dimensional space, it is possible to
reproduce the action of the standard model, with the interpretation
of the Higgs field as a an intermediate ``vector" boson, on a a par
with the photon, $W$ and $Z$ particles. The ``vectoriality" of Higgs
is in the internal space. The gravitational part of the action also
emerges, and all parameters of the model (masses, Yukawa couplings
etc.) are encoded in the zero-dimensional part of the Dirac
operator.

This approach is not without problems, and recently there have
been attempts~\cite{Barrett06, Connes06} to solve some of them by
considering the internal space to have some characteristics of a
\emph{six dimensional} space. The problem is that on the one side
the internal space should be zero dimensional, because it is
represented by a finite algebra, but on the other its chirality
and charge conjugations should behave like those of a space of six
(modulo 8) dimensions. Connes~\cite{Connes06} has solved this by
noting that the concept of deimenion is not uniquely determined in
noncommutative geometry, and considering the KO-dimension,
different from the metric dimension. The two coincide for ordinary
(commutative) manifolds and Dirac operators, but are not
necessarily the same for noncommutative spaces.

In this paper I conjecture that the metric dimension of the
internal (compactified and noncommutative) space is indeed six,
but that all but a finite number of the degrees of freedom are
``frozen" at low energy, smaller than the renormalization scale. I
then go on to further conjecture that this internal space could be
coming from compactified strings. The conjectures here are based
on circumstantial evidence, and no hard calculations are
performed. In particular I do not exhibit the noncommutative
geometry which could give rise to the standard model, but only
indicate some of its features. There are other issues relevant to
the spectral action and the standard model that I do not tackle.
Nevertheless I feel that the considerations presented here,
speculative and unmoulded as they are, can be of interest and
sparkle some interest. A much more ambitious programme would be
the investigation of the possible landscapes of string theory
giving realistic models of particle physics.

The paper is organized as follows. In the first two section I
describe the noncommutative geometry and the standard action
respectively. I then describe in section~\ref{freeze} how a
continuous noncommutative geometry can be frozen to appear finite
dimensional in the context of the standard action. In
section~\ref{internalstring} I speculate how such a mechanism
could be present in a string theory. I conclude with some final
remarks.

\section{The standard model as an almost commutative geometry}
\setcounter{equation}{0}

In this section I give a brief sketch of how the standard model
arises in noncommutative geometry, more details can be found in
the reviews~\cite{G-BMV,Schuckerlisbon}. In the work of Connes and
collaborators~\cite{Connesbook, Connesreal} the standard model is
a particular ``almost commutative" geometry. The functions on this
space are the tensor product of $C(M)$, the algebra of continuous
functions on a four dimensional manifold $M$, times a finite
dimensional algebra which represents an internal, noncommutative,
space. To reproduce the standard model the internal space is taken
to be
\be
{\cal A}_F=\complex\oplus{\mathbb H}\oplus M_3(\complex)\ ,
\ee
where the factors are complex numbers, quaternions and complex
valued three by three matrices. Thus the continuous functions
which describe this noncommutative geometry (and its topology)
form the algebra
\be
{\cal A}=C(M)\otimes {\cal A}_F\ . \label{tripleprodsm}
\ee
The gauge group is given by the unimodular (unitary elements of
the algebra of unit determinant) part of the algebra, and in this
case it correctly reproduces the standard model gauge group, i.e.\
functions valued in $U(1)\times SU(2)\times SU(3)$.

The algebra is represented as operator on the Hilbert space of
spaces, which again is a tensor products of a continuous part
composed of spinors on the four dimensional space, times and
internal space.
\be
{\cal H}={\cal H}_C \otimes {\cal H}_F \label{Hilbert}
\ee
where $C$ is for continuum and $F$ for finite, then it is possible
to split both the spacetime and the internal space spinors into left
and right parts. The internal Hilbert space can be split into the
different chiralities and charge as
\be
{\cal H}_F={\cal H}_L\oplus{\cal H}_R\oplus{\cal H}^c_L\oplus{\cal
H}^c_R
\ee
with $L,R$ referring to chirality, and the the superscript $c$
referring to the charge conjugate (antiparticle) sector. It
comprises all fermionic degrees of freedom of the standard model,
i.e.\ four degrees of freedom for electrons and positron of both
helicities, two for neutrinos, 24 for quarks (including colour),
which makes 30 degrees of freedom per generation, for a total of 90
degrees of freedom. Since spinors already contain the degrees of
freedom of left and right particles and antiparticles there is an
overcounting. Note also that at this stage I am not considering
right-handed neutrinos. I will come back to these aspects. The
representation of ${\cal A}_F$ on ${\cal H}_F$ acts in a different
way in the particle and antiparticle sectors, and this is crucial
for the bimodule structure introduced below. I do not give the
explicit finite dimensional representation, and refer to the
literature for details.

The geometry of (ordinary or noncommutative) spaces is encoded in
the Dirac operator. For example, the (metric) dimension of the space
is given by the exponent in the growth of the number of eigenvalues
of it, i.e., if call the list of $d_k$ these eigenvalues in
ascending order (and repeated according to the degeneracy), then if
there exist two real numbers $c$ and $n$ such that
\be
d_k\le C k^{-n}
\ee
then $n$ is called the metric dimension of the space. If the
operator is finite the metric dimension is of course zero.

For the case at hand also the Dirac operator again is split in the
continuous four dimensional operator and internal, finite matrix,
representing a zero dimensional internal space :
\be
D=\delslash \ \otimes \id + \gamma_5 \otimes D_F\
.\label{definedirac}
\ee
The presence of $\gamma_5$ is conventional, its insertion ensures
that the Laplacian has a similar splitting into an internal and an
external part. The four dimensional part it is the usual
$\delslash=\gamma^\mu\del_\mu$ while $D_F$, in the $L,R$ basis is
\begin{equation}
D_{F} =\small{ \left( \begin{array} {cccc}
0 & {\cal M} & 0 & 0 \\
{\cal M}^{\dag} & 0 & 0 & 0\\
0 & 0 & 0 & {\cal M^*} \\
0 & 0 & {\cal M}^{T} & 0 \end{array} \right)\ .} \label{2.13}
\end{equation}
The $24 \times 21$ mass matrix ${\cal M}$ contain information on the
masses of all fermions, and the Cabibbo-Kobayashi-Maskawa mixing
angles.

The algebra ${\cal A}$, the Hilbert space ${\cal H}$ and the Dirac
operator $D$ form the spectral triple of the standard model. There
are however two other main ingredients to compose a \emph{real}
spectral triple. The first is the chirality operator\footnote{This
operator exists only for \emph{even} spectral triple. Otherwise the
triple is called odd.}
\be
\gamma=\gamma_5\otimes\gamma_F \ ,
\ee
The second is appropriate generalization of the usual charge
conjugation operator ${\cal C}$ to comprise an operator acting on
the finite part.
\be
J={\cal C}\otimes J_F
\ee
The construction of a real spectral triple is very sensitive to
the number of dimensions modulo 8, and the operators $D,J$ and
$\gamma$ have to satisfy some consistency relations:
\bea\label{jdg}
 J^2 &=& \varepsilon \id ~, \nn \\
 J D &=& \varepsilon' D J ~, \nn \\
 J\gamma &=& (-1)^{n/2} \gamma J ~,  \quad \mathrm{if} ~~n ~~
\mathrm{is ~~even},
\eea
and the mod $8$ periodic functions $\varepsilon$ and
$\varepsilon'$ given by
\bea\label{spiche}
 \varepsilon &=& (1,1,-1,-1,-1,-1,1,1)~,  \nonumber\\
 \varepsilon' &=& (1,-1,1,1,1,-1,1,1)~, \qquad n=0,1,\dots,7
\, .
\end{eqnarray}
Furthermore, the map
\be
b \mapsto b^o = Jb^*J^{-1} \, ,
\ee
determines a representation of the opposite algebra ${\cal A}^o$
on ${\cal H}$ which commutes with ${\cal A}$ and its commutators
with $D$
\bea
{} [a, Jb^*J^{-1}] &=& 0~, \label{jcomm} \\
{} [[D,a], Jb^*J^{-1}] &=& 0 ~, ~~~\forall  ~a,b \in {\cal A} ~,
\label{j1order}
\eea
thus giving ${\cal H}$ a \emph{bimodule} structure, whereby $b^o$
effectively ``acts on the right" on the fermionic states.

In this case
\bea
\gamma_F=\begin{pmatrix}
  \sigma_3 & 0 \\
 0 & \sigma_3
\end{pmatrix}\nn\\J_F=\begin{pmatrix}
  0 & \id \\
  \id & 0
\end{pmatrix}
\eea
One can verify that relations~\eqn{jdg} hold with the correct zero
dimensions for a space described by a finite dimensional algebra.

The geometry of the standard model is therefore that of an
\emph{almost commutative} geometry, in the sense that it is the
product of the ordinary  spacetime, times a matrix algebra. So in
the end the structure of spacetime remains unchanged.

\section{The spectral action principle and the standard model}
\setcounter{equation}{0}

I need to write down the action of the model, to be able to build
the standard model. With the Dirac operator one can construct
differential one-forms, which are self-adjoint operators of the form
\be
A=\sum_ia_i[D,b_i] \ \ \ , \ \ a_i,b_i\in{\cal A}
\ee
and a covariant Dirac operator
\be
D_A=D+A+JAJ^\dagger
\ee
Note that the presence of the conjugation by $J$ is crucial for
$D_A$ to act in a symmetric way on particles and antiparticles,
since the representation of ${\cal A}$ threats them in a different
way.

I have now sketched all ingredients for the construction of the
action. This is the sum of bosonic and a fermionic
part~\cite{connes96, chamseddineconnes1, chamseddineconnes2}:
\be
S=\tr\chi\left(\frac{D_A}\Lambda\right)+\bra{\Psi}D_A\ket{\Psi}
\label{spac}
\ee
with $\Psi\in{\cal H}$. Here $\tr$ is the usual trace in the
Hilbert space ${\cal H}$, $\Lambda$ is a \emph{cutoff parameter}
and $\chi$ is a suitable  function which removes all eigenvalues
of $D_A$ larger than $\Lambda$, a smoothened version of a the
characteristic function of the interval [0,1]. The spectral action
has to be read in the Wilson renormalization scheme sense. All
physical quantities are \emph{bare} and they are subject to
\emph{renormalization} thus getting an energy dependent meaning.
The spectral action is invariant under the gauge action of the
inner automorphisms given by
\be
A\to UAU^\dagger+U[D,U^\dagger]
\ee
with $U$ an unitary element od ${\cal A}$.

The computation of the action~(\ref{spac}) is conceptually simple,
but quite involved~\cite{chamseddineconnes2,KastIochSchuckspec}.
One computes the square of the Dirac operator with
Li{\-}chn\'e{\-}ro{\-}wicz' formula~\cite{berline-getzler-vergne}
and the trace with a suitable heat kernel expansion~\cite{Gilkey},
so as to get an expansion in powers of the parameter $\Lambda$.
The key points are general identities valid for differential
elliptic operators of degree $d$ on an $m$-dimensional manifold
$M$ with metric $g$:
\be
\tr(O^{-s})=\frac{1}{\Gamma(s)}\int_0^\infty d t\,
t^{-s}\tr\e^{-tO} \ \ , \ \mbox{Re}(s)\geq 0 \ \ , \label{tracO}
\ee
and (heat kernel expansion)
\be
\tr \e^{-tO}=\sum_{n\geq 0} t^{\frac{n-m}d}\int_M d\mu_g(x)
a_n(x,O) \ .
\ee
The $a_n$'s are called Seeley-de~Witt coefficients and their
expression (as well as the method to calculate them) can be found
in~\cite{Gilkey}. The trace~\eqn{spac} will then depend on the
$a_n$'s and on the ``momenta'' of $\chi$:
\be
\tr\chi(O)=\sum_{n\geq o} f_n a_n(O) \ \ ,
\ee
with
\bea
f_0&=&\int_0^\infty d t\, t\,\chi(t)  \ , \nn\\
f_2&=&\int_0^\infty d t\,\chi(t)  \  , \nn\\
f_{2n+4}&=&\left.(-1)^n\del_t^n\chi(t)\right|_{t=0}  \ .
\eea
If $\chi$ is not too different from the characteristic function of
the interval $[0,1]$ but just a smoothened version of it, it is safe
to assume $f_2=f_4=2f_0=1$. In this case the spectral action just
counts the eigenvalues of $D_A^2$ which are less than the cutoff
$\Lambda^2$.

In~\eqn{spac} appears the covariant Dirac operator, square root the
Laplacian, implicitly it contains information about the metric and
other relevant geometric aspects of the manifold. It should
therefore come as no surprise that they contain Riemann curvature,
Christoffel symbols etc. What is more surprising is that one obtains
an action which contains all of the ingredients of the standard
model \emph{plus} gravity. The calculation is quite long and
involved, for details refer to~\cite{chamseddineconnes2,
KastIochSchuckspec, Schuckerlisbon}. The final result is:
\be
S_B =\int_M d \mu_g \left(I_1\Lambda^2+I_2+I_3\frac1{\Lambda^2}+
O\left(\frac1{\Lambda^4}\right)\right)\ .
\ee
The first coefficient is constant:
\be
I_1=\frac{45}{8\pi^2} \ ,
\ee
and gives as leading term a cosmological constant. This
coefficient is subject to the renormalization group flow, and to
agree with the observed absence (or at least smallness) of the
cosmological constant, a high degree of fine tuning is necessary.
The second coefficient is
\be
I_2=\frac 1{16\pi^2}\left(-15 R -8 K_1 |\Phi|^2\right)\ ,
\ee
with $R$ the Ricci scalar, $K_1=\tr (3M^\dagger_uM_u+
3M^\dagger_dM_d+M^\dagger_eM_e)$, the $M$'s are mass matrices for
up and down quark-types, and leptons. This term contains the
Einstein-Hilbert Lagrangian as well as the quadratic term of the
Higgs potential. The remaining terms of the potential, as well as
the kinetic terms for the gauge fields are in the term of order
$\Lambda^{-2}$:
\bea
I_3&=& \frac 1{16\pi^2}\left(240 F^{\mu\nu}F_{\mu\nu}+
12G^{\mu\nu}G_{\mu\nu}+4K_1|{D_A}_\mu\Phi|^2 \right. \nn\\
&&\left.  -\frac 23 K_1 R |\Phi|^2 + K_2|\Phi|^4 - \frac 94
C^2\right) +\mbox{surface terms}
\eea
with $F$ and $G$ the gauge field strength for the abelian and
nonabelian part respectively, $K_2=\tr (3(M^\dagger_uM_u)^2+
3(M^\dagger_dM_d)^2+(M^\dagger_eM_e)^2)$, and $C$ the Weyl tensor
defined as
\be
C_{\mu\nu\sigma\rho}=R_{\mu\nu\sigma\rho}-g_{\nu[\rho}R_{\mu|\sigma]}
+\frac 16(G_{\mu\rho}g_{\nu\sigma}-g_{\mu\sigma}g_{\nu\rho})R\ .
\ee

Hence the full action reproduces all terms of the standard model and
gravity, plus some unavoidable terms quadratic in the curvature. As
I said the action must be read in a Wilsonian sense, and to make
predictions it is necessary to study the renormalization of the
coupling constants. The model sketched here has relations among
coupling constants which are $g_3=g_2=(5/3)^{1/2} g_1$, a large
renormalization is needed and one
gets~\cite{Schuckerlisbon,SchuckerKnecht}, for $\Lambda\simeq
10^{13-17}$ GeV the mass $m_H= 175.1 ^{+5.8}_{-9.8}$ GeV. The value
for $m_H$ is (at present) still realistic, while the relation among
couplings is not.

The spectral action principle gives an interpretation of the
standard model which shows a path to a deeper comprehension of it.
In this sense the mass relations, or other possible phenomenological
predictions are not its nicest feature. What is most important is
the fact that Connes (and collaborators) succeeded in putting a
generic Yang-Mills theory in a larger geometrical context, so that
the model is at the same time more rigid, and more flexible. It is
more rigid in the sense that not all gauge groups, and all
representations, can be a candidate. In fact the noncommutative
geometry model building is severely restricted. It is important that
the Standard Model can be seen in this framework, while the (non
supersymmetric) $SU(5)$ unified theory, ruled out by experiments,
cannot. It is more flexible in that it sees the standard model as
coming from an internal (noncommutative) space. This space does not
have to be an ordinary manifold (as in Kaluza-Klein), and so this
opens the way for several generalizations. If this approach is
correct phenomenological and experimental predictions will certainly
follow.

There are however several problems with this approach. A partial
list goes as follows:

\begin{enumerate}

\item {\sc The model is Euclidean \label{lorentz}}

The whole construction of spectral triples is mathematically
consistent only for compact Euclidean manifold. The non compact
case can be dealt without excessive problems (see for
example~\cite{Selene}) but putting a Lorentzian signature is less
natural . The wave functions of the fields are not anymore a
Hilbert space, just to mention one problem. There have been
however constructions of Lorentzian spectral triples~\cite{G-BMV,
Strohmaier, vanSuijlekom}.

\item{\sc Fermion doubling \label{doubling}}

I have already noted this. The problem~\cite{doubling} is that the
four dimensional part of the spectral triple requires the presence
of fields which are ordinary Dirac fermions, but the constructions
put in the finite internal Hilbert space all degrees of freedom, so
that the four degrees of freedom corresponding which comprise a
Dirac spinor field (say electron and positron of right and left
chiralities) are multiplied tensorially by the same degrees of
freedom appearing in the internal Hilbert space.

\item{\sc Neutrino masses \label{neutrinos}}

At the time of  construction of the original model neutrinos where
believed to be massless, so the model was built this way. There is
now experimental evidence to the contrary, so the model has to
accomodate these masses, either as Dirac or Majorana
masses~\cite{Schelp,BarrettMartins}. Majorana masses create problems
with the \emph{first order condition}~\eqn{j1order} which imposes
that forms have to commute with functions. The presence of a
Majorana mass introduces in the matrix of the finite part of the
Dirac operator which connect the left with the right chirality, and
this spoils the commutation relation $[JaJ^{-1},[D,b]]$. Dirac
masses are instead spoil Poincar\'e duality, the noncommutative
geometry version of the duality between forms of degree $n$ and of
degree $d-n$~\cite{Connesreal}. While a violation of the first order
condition creates unsurmountable problems with gauge
invariance\footnote{In principle it is possible that the first order
condition is satisfied up to an infinitesimal, this happens
in~\cite{DLPS}.}, it is not clear which problems a violation of
Poincar\'e duality would create. Phenomenologist in general favour
strongly the see-saw mechanism to give mass to the neutrinos, and
this requires necessarily Majorana masses.

\item{\sc The model is \emph{ad hoc} \label{adhoc}}

The construction is very complicated, expecially the
representation of the Hilbert space. Everything is put there by
hand, just to fit the standard model. While it is true that the
aim of the game (at this stage) is to \emph{understand} the
standard model, not predict it, nevertheless it would be desirable
to have some of the features come out in a more natural way.

\item{\sc Classical vs. Quantum}

The original Connes-Lott is classical, and its mass relations do
not survive quantization~\cite{AG-BM}, although some relations can
still be inferred~\cite{fuzzymass}. The spectral action on the
other side is quantum by nature, as it has to read in the Wilson
renormalization scheme, but suffers from the fact that there are
relations among coupling constants, which are the same as in
SU(5), and are not phenomenologically favoured. In general one
would like some quantization process more on a par with the rest
of the mathematical construction. In this respect more work on the
roots of the spectral action is necessary.

\item{\sc Fine Tuning}

The spectral action requires several instances of fine tuning, to
start with a huge cosmological constant. An analysis of this
problem in the light of the recent indications of a non-vanishing
cosmological constant would be of great interest.

\item{\sc The model is almost commutative}

In other words this is an effective model, and one would like to
see more of the full power of noncommutative geometry at work,
rather than just some discrete Kaluza-Klein. For instance, where
is the internal (matrix) space coming from?

\end{enumerate}

Some of these drawbacks have been examined in two recent articles.
Let us examine them in turn. In ~\cite{Barrett06} Barret deals
with points~\ref{lorentz} and~\ref{doubling}. Let us recall that
the problem with fermion doubling pointed out in~\cite{doubling}
is not so much the presence of more states than one would expect.
This can be fixed projecting out the unwanted states, and
identifying the remaining ones. The problem is in that the
projection spoils the action. If the Hilbert space is the product
of the continuum part by the discrete part indicated
in~\eqn{Hilbert}.

The troublesome terms in ${\cal H}$ are the ones coming form the
crossed products of states with different chiralities. These do
not properly transform under CP. The idea in~\cite{doubling} was
to project out those states. This is possible, and the fermionic
action can be correctly written to contain only the proper degrees
of freedom. But the problem is that if one uses a reduced Hilbert
space, then the trace which gives the bosonic action will be over
a reduce space, and the final result is not the correction for the
standard model, both for the Connes-Lott model or the spectral
action. We tried to fix this by giving a physical meaning to the
extra ``mirror" fermions~\cite{mirror}. This is not a satisfactory
solution which somehow nullify the very interesting connections
between the quantum number assignments of the standard model and
cancellation of anomalies. Moreover there is a fine tuning
problem. In~\cite{G-BIS} a slightly different form of the bosonic
part of the Connes-Lott action, reinstating the missing pieces of
the action. The problem here is that in the commutative limit this
action vanishes, and moreover no solution for the spectral action
is offered .

Barrett notices that if one uses the chirality $\gamma$ and $J$
operators coming from the Minkowski, and not Euclidean, assignments
for the four-dimensional part, and the operators for a
\emph{six-dimensional} space for the internal part, then the
doubling disappears from the Fermionic part of the action if one
considers as Hilbert space the subspace of ${\cal H}$ which is
eigenspace of $\gamma$and $J$, with eigenvalue $\ii$ and 1
respectively. Barret does not consider the bosonic part.

This is similar to the projection in~\cite{doubling}, in fact the
proper projection operator is $P=\frac{1\pm\gamma}2$ on the
particle and antiparticle sectors. The problem is that with this
assignment there is no way to have the bosonic action come out
correct. In fact the calculation is similar the one done
in~\cite{mirror}, without the mirror fermions and with a slight
modification of the Dirac operator.

At the same time as the paper by Barrett, a paper by
Connes~\cite{Connes06}  appeared. Like Barrett, Connes notices the
mismatch between the dimension of the internal space (zero), and the
values of $\varepsilon$ and $\varepsilon'$ required, which would be
the ones indicated in~\eqn{spiche} by dimension 6 (mod.~8). This is
interpreted as a mismatch between the \emph{metric} dimension, and
the {\sl KO}-dimension. The former is the one encoded in the growth
of the eigenvalues of the Dirac operator. The latter is the one
encoded in~\eqn{spiche}. For ordinary, commutative, manifolds these
definitions coincide, but this may not be so for noncommutative
geometries. It is known that (even ordinary) spaces can have
different definitions of dimension which do not necessarily
coincide. This is common for fractals. For noncommutative geometries
this happens commonly for spaces such as the Podles
sphere~\cite{Podles} which have metric dimension zero, but
KO-dimension 2~\cite{DLPS}.

There also in Connes' paper the assignments for $J_F$ and
$\gamma_f$ are changed to make them like those a six-dimensional
space. It turns out that this is equivalent (in the Euclidean
case) to just reversing the sign of $J$ and $\gamma$ in the
antiparticle sector. Since he is in the Euclidean framework the
total KO-dimension is now $10=2=26$ (mod 8), and the connection
with strings immediately comes to mind. There is another subtle,
but not minor, change in this version of the spectral action,
namely the presence of $J$ in the fermionic action, so that now
the action reads
\be
S'=\tr\chi\left(\frac{D_A}\Lambda\right)+\bra{J\Psi}D_A\ket{\Psi}
\label{spac'}
\ee
with $\Psi$ belonging to the projected Hilbert space of physical
particles. The presence of $J$ is crucial for the correct pairings
of particles and antiparticles. What is left is then to crank
again the machine, which in Connes' word is an ``excruciating
calculation, rewarded by the result of an action which reproduces
the gauge sector of the standard model. The novel choice of $J$
and $\gamma$, and the modification of the fermionic action
eliminate the doubling and also allows for the presence of
right-handed neutrinos and the see-saw mechanism.

Therefore the internal space of noncommutative geometry has metric
dimension zero,and {\sl KO}-dimension six. In the next section we
will see how a (noncommutative) six dimensional internal space can
give rise, with the spectral action, to a noncommutative geometry
which is well approximated by an almost commutative one.

\section{Nonabelian Yang-Mills from internal noncommutative
spaces \label{freeze}} \setcounter{equation}{0}

The spectral action is a natural evolution of the central philosophy
on noncommutative geometry, for which the geometry is given by the
Dirac operator, it is however an effective action, and for example
there is a nonperturbative versions of the action~\cite{zappafrank}
based on superconnection~\cite{Quillen}. Being an effective action,
valid only at certain scales, it is possible that the geometry we
are probing with the standard model is only a low energy geometry.
In this section I see how an effective geometry, with the
characteristics of the standard model, might emerge.

The bosonic part of the spectral action is sensitive only to the
eigenvalues of the Dirac operator which are less than the energy
scale given by $\Lambda$. This is the renormalization scale, a
quantity that is safe to assume of an order of magnitude between the
grand unification scale ($\sim10^{16}$~GeV) and Planck energy
($\sim10^{19}$~GeV). As for the spacetime part, even considering
(for technical or physical reasons) spacetime compact, there are
obviously many eigenvalues much smaller than $\Lambda$, and this
requires the use of heath kernel techniques.

If however the internal space is compactified on a scale of the
order of $\Lambda$ things are different. Consider again, as
in~\eqn{tripleprodsm} a manifold which is the product of spacetime

by an internal space, which is taken to be a noncommutative space
described by a noncommutative algebra ${A_I}$, represented on an
Hilbert space ${H_I}$. One may think of the noncommutative space as
a deformation of a compact ordinary manifold. So the full algebra
and Hilbert spaces of the triple will be
\bea
{\cal A}&=&C(M)\otimes {\cal A}_I\ , \nn\\
{\cal H}&=&{\cal H}_C \otimes {\cal H}_I \nn\\
\eea

Consider now the case in which the eigenvalues of the Dirac
operator, call them $d_i$, have the characteristic that not only few
of them are small (with respect to $\Lambda$), but that also their
difference is small only for a limited range of indices
\be
|d_k-d_l| \gg\Lambda\ \mbox{if}\ k,l > N \label{gerarchia1}
\ee
i.e.\ only a finite number of pairs of eigenvalues are much smaller
than the renormalization scale. The details of the noncommutative
space are not important at this stage. What is important is that,
since $A_I$ is a $C^*$-algebra, it can always be represented as
bounded operator on ${\cal H}_I$. Therefore it is possible to
consider ${\cal A}_I$ to be made of infinite matrices, and consider
them in the basis which diagonalizes ${D_I}_0$. An element of ${\cal
A}$ will be denoted by $a_{kl}(x)$. A generic one-form will be of
the kind:
\be
A_{kl}(x)=\sum_i a^{(i)}[D_0,b^{(i)}]=\sum_i\sum_{m\in \zed}
a^{(i)}_{km}(x)(d_l-d_m)\delslash b^{(i)}_{ml}(x)
\ee
I am considering low energy fluctuations on the vacuum, and
therefore $a$and $b$ are much smaller than $\lambda$, so
that\footnote{I am not considering $J$ for the moment, to keep
things simple.} $D_A = (\delslash+d_k) \delta_{kl}+A_{kl}$ has all
elements, except the first $N$ rows and columns, much larger than
$\Lambda$. Because of the function $\chi$, then effectively only the
$N\times N$ minor contributes to the action, which therefore becomes
effectively the spectral action for an almost commutative geometry
with gauge group $U(N)$.

With the same mechanism, and different scales, it is possible to
have scenarios for which the internal space is a product of matrix
algebras. It will suffice to modify \eqn{gerarchia1} into
\be
|d_k-d_l| \gg\Lambda\ \mbox{if}\ k,l > N_1, \ \mbox{or\ }
N_1<k,l<N_2 \ldots \label{gerarchia2}
\ee
The important point is that the spectral action measures the
\emph{fluctuations} of the space, as long as they do not exceed the
scale $\Lambda$, and this may happen only if there are no quantities
which exceed this scale, if the the theory can be considered
perturbatively.

Let us try to be slightly more precise (but still highly
speculative). Consider the internal space to be a deformation of a
\emph{two-dimensional} space. I assume the deformation to be so
``mild" that the relevant structures of the undeformed case survive.
The internal Hilbert space in this case is made of two dimensional
spinors, whose components are functions on the noncommutative space.
The other elements of internal part of the spectral triple can be
\be
{D_I}_0=\begin{pmatrix}
  0 & \nabla_2+\ii\nabla_i \\
  \nabla_2-\ii\nabla_i & 0
\end{pmatrix}\ , \ \gamma_I=\begin{pmatrix}
  \id & 0 \\
  0 & \id
\end{pmatrix}\ , \ J_I=\begin{pmatrix}
  0 & -C \\
  C & 0
\end{pmatrix}\ , \
\ee
With $C$ complex conjugation, $\nabla_i$ the two covariant
derivatives which take into account the curvature of the space. We
are making lots of assumptions here, and there is no warranty that
these operators satisfy relations~\eqn{jdg} for a two-dimensional
space, nor that they exist for that matter.

The eigenvalues of ${D_I}_0$ come in pairs differentiated only by
the sign. The process of diagonalization of ${D_I}_0$ can be done in
two steps, first diagonalize $\nabla_2+\ii\nabla_1$, or rather its
deformed equivalent, with the infinite dimensional matrix $u$, hence
\be
{D'_I}_0 =
\begin{pmatrix}u&0\\0&u\end{pmatrix} {D_I}_0
\begin{pmatrix}u^\dagger&0\\0&u^\dagger\end{pmatrix}
=
\begin{pmatrix}0&d\\d^\dagger&0\end{pmatrix}
\ee
where $d$ is diagonal. Then block diagonalize with the unitary
matrix
\be
U=\frac{1}{\sqrt{2}}\begin{pmatrix}
  \id & \id \\
  \tilde d^* & -\tilde d^*
\end{pmatrix}
\ee
where $\tilde d$ is a diagonal matrix of elements $d_k/|d_k|$ if
$d_k\neq 0$, 1 otherwise, hence all of its elements are phases. In
the new basis
\bea
U\gamma_IU^{-1}&=&\begin{pmatrix}
  0 & -\id \\
  -\id & 0
\end{pmatrix}\nn\\
UJ_IU^{-1}&=&\begin{pmatrix}
  \tilde d C & 0 \\
  0 & -\tilde d C
\end{pmatrix}
\eea
If now the we operate the reduction to the minor of small matrix
elements of $D_A$, we pick up a minor composed of the first few
rows and columns of both blocks and we have a duplication of the
internal structure, inherited from the two-dimensional spinors.
The chirality $\gamma_I$ maintains the same structure in the
minor, and the phases appearing in $J_I$ can be eliminated with a
further conjugation by a diagonal unitary matrix.

Effectively only a matrix part of the noncommutative algebra ${\cal
A}_I$ contributes to the spectral action. We end up with a
$U(n)\times U(n)$ gauge theory, of the kind very familiar in
noncommutative geometry~\cite{ConnesLott, Connesreal}. With a
judicious choice of the Dirac operator the gauge group can be broken
to a smaller diagonal group. A crucial difference is that however
the commutation relation of the reduced $D_I, J_I$ and $\gamma_I$
have remained those of the original two-dimensional theory, and are
block reduction of those of a zero-dimensional space. The point is
that in reality the internal space is two-dimensional, but the
particular choice of eigenvalues effectively \emph{freezes} out the
higher modes.

\section{Noncommutative internal space and strings \label{internalstring}}
\setcounter{equation}{0}

In this section I will attempt a connection between the spectral
action and string theory, arguing on the possibility of a way to
extract the standard model from a string theory. I will discuss how
the internal space of strings could give rise, with a freezing
process like the one discussed earlier, to a model with the
characteristics of the standard model in the framework of the
spectral action.

The first step is the construction of the noncommutative geometry
of strings. Connections between noncommutative geometry and
strings have become common in the literature since the work of
Seiberg and Witten~\cite{SeibergWitten}, but attempts to construct
the noncommutative geometry of strings, in the framework of
Connes' spectral geometry, go back to Frohlich and Gawedzky
\cite{FrohlichGawedzky} and has been further developed in
\cite{FGR, LizziSzaboprl, LizziSzabodual, LLSstring}.

I now give the essential elements of the construction of the
(closed) strings spectral triple, the Frohlich-Gawedzky triple, and
refer to the above papers, expecially~\cite{LizziSzabodual,
LLSstring}, for details. Consider bosonic strings compactified on a
$d$ dimensional torus ${\mathbb T}_d$ with radii $L_i$. This torus
(or a deformation of it) will be the internal space. The basic
operators of the theory are the Fubini-Veneziano fields
\be
X_\pm^i(z_\pm)=x_\pm^i+\ii g^{ij}p_j^\pm\log
z_\pm+\sum_{n\neq0}\frac1{in}\,\alpha_n^{(\pm)i}z_\pm^{-n}
\label{FVfieldsdef}
\ee
and the chiral Heisenberg fields
\beq
\alpha_\pm^i(z_\pm)=-\ii\partial_{z_\pm}X_\pm^i(z_\pm)=
\sum_{n=-\infty}^\infty
\alpha_n^{(\pm)i}z_\pm^{-n-1}~~~~~~;~~~~~~\alpha_0^{(\pm)i}\equiv
g^{ij}p_j^\pm \label{heisfieldsdef}\eeq where
$z_\pm\in\complex\cup\{\infty\}$. The $z^\pm$'s are the world sheet
coordinates and $g^{ij}$ is the target space metric. The center of
mass of the string is $x=x_++x_-$, its momentum $p=p_++p_-$, the
winding is $p=p_+-p_-$. The $\alpha_n$ for $n$ negative (resp.\
positive) are the creation (resp.\ annihilation) modes for the
oscillatory modes of the string.

The states of this string are described by an Hilbert space which
takes into account the momentum, the winding number and the left
and right  modes of the string:
\be
{\cal H}_I= L^2_S\left({\mathbb T}_d\times {\mathbb T}_d^*\right)
\otimes{\cal F}^+\otimes{\cal F}^- \label{hilbertdef}
\ee
where by $L^2_S$ I mean the square integrable spinors, ${\cal
F}^\pm$ are the Fock spaces generated by the left and right moving
oscillatory creation operators, ${\mathbb T}_d^*$ is the dual torus
whose presence is due to the states in which the strings winds
around the compact dimension. The radii of the dual torus are
$L_i^{-1}$. The states of the string are created by the action of
\emph{vertex operators} acting on a vacuum (operator states):
\be
\ket{\Psi}=V_\Psi(z_+,z_-)\ket{0}
\ee
The vertex operator algebra is specific of a particular background
and specific string theory, it is natural to take as second
element of the spectral triple the (properly regularized) vertex
operator algebra of the theory. For a toroidal background the
fundamental vertex operators for the low lying states (tachyonic
for the bosonic string) are of the kind
\be
V_{q^\pm}=:\e^{\ii q^\pm X_\pm}: \label{tachyons}
\ee
where $:\cdot:$ means normal ordering with respect to the creation
and annihilation operators and the $q$'s are the momenta of the
states.

The left and right movers of the theory split, so that each copy
has its set of reciprocally anticommuting gamma matrices
\be
\left\{\gamma^\pm_i,\gamma_j^\pm\right\}=\pm\,2g_{ij}
\ee
each sector has his own $\gamma^\pm=\gamma^\pm_1\ldots\gamma^\pm_d$,
and I define
\be
\gamma_I=(-1)^d\gamma^+\gamma^-
\ee
Each sector has his own Dirac operator as well:
\be
D^\pm=\gamma^\pm_\mu z_\pm\alpha_\pm^\mu(z_\pm)
\ee
with which I can build two Dirac operators
\be
D_I=D^++D^-\ \ ,\ \ \bar D_I=D^+-D^-
\ee
The two operators are unitarily related by \emph{T-duality
transformation}. This transformation exchanges winding and
momentum modes. Ignoring the oscillatory modes, and with the usual
identification of the momentum operator with the derivative, the
operator $D_I$ becomes the usual Dirac operator on the
compactified space, while $\bar D_I$ is the operator on the dual
torus. This can be identified with the set of ``position" for the
dual (in the sense of position-momentum duality) of the winding
number.

The last element is $J_I$, which can be taken to act on the
Hilbert space as:
\be
J_I\ket{\Psi}=V_\Psi^\dagger\ket{\Psi}
\ee
The bimodule structure is preserved, in fact:
$$
V_1J_IV_2J_I\ket{\Psi}=V_1J_IV_2J_IV_\Psi\ket{0}=
V_1J_IV_2V_\Psi^\dagger\ket{0}= V_1V_\Psi V_2^\dagger\ket{0}
$$
\be
J_IV_2J_IV_1\ket{\Psi}=J_IV_2J_IV_1V_\Psi\ket{0}=J_IV_2V_\Psi^\dagger
V_1^\dagger\ket{0}= V_1V_\Psi V_2^\dagger\ket{0}
\ee
The relations between $D_I, \gamma_I$ e $J_I$ satisfy those of a
$2d$-dimensional space. The reasons for this doubling of the
dimensions is that both momentum and winding modes contribute. In
fact we have shown in~\cite{LLSstring} that there is a range of
energies for which the internal space is effectively composed of two
copies of a noncommutative torus. We have also
shown~\cite{LizziSzaboprl} that on the contrary, if the
compactification is very large then the degrees of freedom are those
of commutative ``subtriples" which reproduces spacetime, with the
winding modes playing, as it should, no role.

Let us now, very heuristically, discuss the features of a string
internal landscape which could give rise to the standard model.
Consider the space made of four uncompactified dimensions, for which
the usual geometry will apply, and some dimensions compactified on a
small scale. I now speculate how some internal space of strings
could give rise to an almost commutative geometry with some features
of the standard model. Assume that the scale of compactification of
the internal space, and of the lowest eigenvalues of the Dirac
operator) will be of order $\Lambda$, while the scale of the
oscillatory modes is higher, at the Planck scale. This is not a new
scenario in string theory~\cite{AA-HDD}. This means however that we
can ignore the oscillatory modes in a spectral action, but that we
would have to take into account both the momentum and winding modes.

A torus does not give the desired hierarchy of eigenvalues, so we
suppose that the compactified geometry is deformed to give the
desired set of eigenvalues, but still keeps the main features of the
toroidal case. In particular the presence of momentum and winding
number modes. One expects that, as in the torus case, the
noncommutativity in the internal space comes form the interplay of
these two modes, but if the space is not a torus the two spaces are
not the same.

Let us look at the features that the noncommutative geometry of
compactified strings has in common with the standard model. There
are two sectors, the left and right movers. They are the eigenspaces
of the chirality operator $\gamma$. The two sectors are connected by
T-duality, which exchanges momentum and winding modes. Again the
torus is not apt to reproduce the standard model, since it has a
perfect symmetry between these two sector, in contrast with the
chiral characteristics of the standard model, but is should be
possible to envisage a (deformed) geometry for which the momenta
have a different symmetry from the windings. The charge conjugation
operator acts transforming the string into the hermitian conjugate,
exchanging left and right sectors, and giving the bimodule
structure. The Dirac operator feels the geometrical structure of the
internal space, which is not necessarily a deformation of a
commutative space, but just an algebra generated by a vertex
operator algebra. The value of its eigenvalues can therefore be
quite general.

The low lying states which contribute to the action are a vacuum
state, and then the $d$ states created by the
operators~\eqn{tachyons} acting on it. For $d=3$, in view of the
argument described in section \ref{freeze}, one would have that the
relevant states of the vertex algebra freezes to a $\complex\oplus
M_3(\complex)$. The left-right structure we conjecture is given by
T-duality. A chiral theory would be a consequence of a non T-dual
state. At present I have do not see how to obtain a quaternionic
structure, nor the presence of more than one generation.

\section{Final Remarks}
\setcounter{equation}{0}

The speculative character of this paper is such that it poses many
questions and gives no answers. The main task is to build a geometry
with at least some of the desired features of the theory, in terms
of the spectrum of the Dirac operator. On one side this is a
difficult task because there are not so many noncommutative
manifolds which are studied sufficiently. On the other side the
possibility to have noncommutative space gives added degrees of
freedom which effectively enlarge the scope of the investigation.
What I envisage is a programme ``top-down", which starting form the
internal geometries sees which sort of gauge/gravity model they can
give. This is dual to the conventional approaches in noncommutative
geometry which start from the standard model (with all of its
parameters) and try to fit the geometry which fits it.

String theory provides us with an extreme wealth of possible
internal geometries, and structure which I do not have considered
here, supersymmetry and branes the most important ones, they will
definitely have consequences for the matter discussed here. In
general more work is needed (independently of the conjectures of
this paper) to understand string theory in the framework of the
spectral action.

\subsection*{Acknowledgments}
I would like to thank G.~Esposito, J.M.~Gracia-Bond\'\i a, G.~Landi
and R.~Szabo for discussions and correspondence. This work has been
supported in part by the {\sl Progetto di Ricerca di Interesse
Nazionale {\em SInteSi}}. I would like to thank S.~Clement-Lorenzo
for hospitality in Ixelles-Elsene, where this whole work has been
performed.


\begin{thebibliography}{99}

\baselineskip=12pt
\bibitem{ConnesLott}
A.~Connes and J.~Lott,
  ``Particle models and noncommutative geometry (expanded version),''
  Nucl.\ Phys.\ Proc.\ Suppl.\  {\bf 18B}, 29 (1991).
  %%CITATION = NUPHZ,18B,29;%%

\bibitem{Connesbook} A. Connes, {\it Noncommutative Geometry}, Academic Press,
 (1994).

\bibitem{ticos} J.M.~Gracia-Bondia, J.C.~Varilly, H.~Figueroa, {\it Elements of
noncommutative geometry}, Birkhauser, 2000.

\bibitem{GreenSchwarzWitten}
 M.B. Green, J.H. Schwarz and E. Witten, {\it Superstring Theory} (Cambridge University Press, 1987).


\bibitem{Polchinskibook}
J. Polchinski, {\it String Theory} (Cambridge University Press,
1998).


\bibitem{chamseddineconnes1}
A.~H.~Chamseddine and A.~Connes,
  ``Universal formula for noncommutative geometry actions: Unification of
  gravity and the standard model,''
  Phys.\ Rev.\ Lett.\  {\bf 77}, 4868 (1996).
  %%CITATION = PRLTA,77,4868;%%

\bibitem{chamseddineconnes2}
A.~H.~Chamseddine and A.~Connes,
  %``The spectral action principle,''
  Commun.\ Math.\ Phys.\  {\bf 186}, 731 (1997)
  [arXiv:hep-th/9606001].
  %%CITATION = HEP-TH 9606001;%%


\bibitem{Barrett06}
J.~W.~Barrett,
   ``A Lorentzian version of the non-commutative geometry of the standard model
  of particle physics,''
  arXiv:hep-th/0608221.
  %%CITATION = HEP-TH 0608221;%%



\bibitem{Connes06}
A.~Connes,
  ``Noncommutative geometry and the standard model with neutrino mixing,''
  arXiv:hep-th/0608226.
  %%CITATION = HEP-TH 0608226;%%

\bibitem{G-BMV}
C.~P.~Martin, J.~M.~Gracia-Bondia and J.~C.~Varilly,
  ``The standard model as a noncommutative geometry: The low-energy regime,''
  Phys.\ Rept.\  {\bf 294}, 363 (1998)
  [arXiv:hep-th/9605001].
  %%CITATION = HEP-TH 9605001;%%


\bibitem{Schuckerlisbon}
T.~Sch\"ucker, ``Geometries and forces,'' talk given at European
Mathematical Society Summer School on Noncommutative Geometry and
Applications, Monsaraz and Lisbon, Portugal, 1-10 Sep 1997.
  [arXiv:hep-th/9712095].
  %%CITATION = HEP-TH 9712095;%%

\bibitem{Connesreal}
A.~Connes,
  ``Noncommutative Geometry And Reality,''
  J.\ Math.\ Phys.\  {\bf 36}, 6194 (1995).
  %%CITATION = JMAPA,36,6194;%%

\bibitem{connes96}
A.~Connes,
  ``Gravity coupled with matter and the foundation of non-commutative
  geometry,''
  Commun.\ Math.\ Phys.\  {\bf 182}, 155 (1996)
  [arXiv:hep-th/9603053].
  %%CITATION = HEP-TH 9603053;%%

\bibitem{KastIochSchuckspec}
   B.~Iochum, D.~Kastler and T.~Sch\"ucker,
  ``On the universal Chamseddine-Connes action. I: Details of the action
  computation,''
  J.\ Math.\ Phys.\  {\bf 38}, 4929 (1997)
  [arXiv:hep-th/9607158].
  %%CITATION = HEP-TH 9607158;%%

\bibitem{berline-getzler-vergne}
N. Berline, E. Getzler, M. Vergne, {\it Heat Kernels and Dirac
Operators} (Springer-Verlag, 1991).

\bibitem{Gilkey}
P.B. Gilkey, {\it Invariance Theory, The Heat Equation, And the
Atiyah-Singer Index Theorem}, 2nd edition, Studies in Advanced
Mathematics (CRC Press, 1995).

\bibitem{SchuckerKnecht}
M.~Knecht and T.~Schucker,
  %``Spectral action and big desert,''
  Phys.\ Lett.\ B {\bf 640} (2006) 272
  [arXiv:hep-ph/0605166].
  %%CITATION = HEP-PH 0605166;%%

\bibitem{Selene}
J.~M.~Gracia-Bondia, F.~Lizzi, G.~Marmo and P.~Vitale,
  %``Infinitely many star products to play with,''
  JHEP {\bf 0204} (2002) 026
  [arXiv:hep-th/0112092].
  %%CITATION = HEP-TH 0112092;%%

\bibitem{Strohmaier}
A.~Strohmaier,
  %``On Noncommutative and semi-Riemannian Geometry,''
  J.\ Geom.\ Phys.\  {\bf 56} (2006) 175
  [arXiv:math-ph/0110001].
  %%CITATION = MATH-PH 0110001;%%

\bibitem{vanSuijlekom}
  W.~D.~van Suijlekom,
  ``The noncommutative Lorentzian cylinder as an isospectral deformation,''
  J.\ Math.\ Phys.\  {\bf 45} (2004) 537
  [arXiv:math-ph/0310009].
  %%CITATION = MATH-PH 0310009;%%


\bibitem{doubling}
F.~Lizzi, G.~Mangano, G.~Miele and G.~Sparano,
  ``Fermion Hilbert space and fermion doubling in the noncommutative  geometry
  approach to gauge theories,''
  Phys.\ Rev.\ D {\bf 55}, 6357 (1997), [arXiv:hep-th/9610035].
  %%CITATION = HEP-TH 9610035;%%

\bibitem{Schelp}
  R.~Schelp,
  %``Fermion masses in noncommutative geometry,''
  Int.\ J.\ Mod.\ Phys.\ B {\bf 14}, 2477 (2000)
  [arXiv:hep-th/9905047].
  %%CITATION = HEP-TH 9905047;%%

\bibitem{BarrettMartins}
J.~W.~Barrett and R.~A.~Dawe Martins,
  ``Non-commutative geometry and the standard model vacuum,''
  J.\ Math.\ Phys.\  {\bf 47} (2006) 052305
  [arXiv:hep-th/0601192].
  %%CITATION = HEP-TH 0601192;%%


\bibitem{DLPS}
L.~Dabrowski, G.~Landi, M.~Paschke and A.~Sitarz,
  ``The Spectral Geometry of the Equatorial Podles Sphere,''
  arXiv:math.qa/0408034.
  %%CITATION = MATH-QA 0408034;%%



\bibitem{AG-BM}
E.~Alvarez, J.~M.~Gracia-Bondia and C.~P.~Martin,
  ``Anomaly Cancellation And The Gauge Group Of The Standard Model In Ncg,''
  Phys.\ Lett.\ B {\bf 364} (1995) 33
  [arXiv:hep-th/9506115].
  %%CITATION = HEP-TH 9506115;%%

\bibitem{fuzzymass}
B.~Iochum, D.~Kastler and T.~Sch\"ucker,
  ``Fuzzy Mass Relations In The Standard Model,''
  [arXiv:hep-th/9507150].
  %%CITATION = HEP-TH 9507150;%%

\bibitem{mirror}
  F.~Lizzi, G.~Mangano, G.~Miele and G.~Sparano,
  ``Mirror fermions in noncommutative geometry,''
  Mod.\ Phys.\ Lett.\ A {\bf 13}, 231 (1998)
  [arXiv:hep-th/9704184].
  %%CITATION = HEP-TH 9704184;%%

\bibitem{G-BIS}
J.~M.~Gracia-Bondia, B.~Iochum and T.~Schucker,
  ``The standard model in noncommutative geometry and fermion
doubling,''
  Phys.\ Lett.\ B {\bf 416}, 123 (1998)
  [arXiv:hep-th/9709145].
  %%CITATION = HEP-TH 9709145;%%

\bibitem{Podles}
P.~Podles,
  ``Quantum spheres,''
  Lett.\ Math.\ Phys.\  {\bf 14} (1987) 193.
  %%CITATION = LMPHD,14,193;%%

\bibitem{zappafrank}
 H.~Figueroa, J.~M.~Gracia-Bondia, F.~Lizzi and J.~C.~Varilly,
   ``A nonperturbative form of the spectral action principle in  noncommutative
  geometry,''
  J.\ Geom.\ Phys.\  {\bf 26}, 329 (1998)
  [arXiv:hep-th/9701179].
  %%CITATION = HEP-TH 9701179;%%

\bibitem{Quillen}
D. Quillen, ``Superconnections and the Chern character," Topology
{\bf 24}(1985) {89}.

\bibitem{SeibergWitten}
 N.~Seiberg and E.~Witten,
  ``String theory and noncommutative geometry,''
  JHEP {\bf 9909}, 032 (1999)
  [arXiv:hep-th/9908142].
  %%CITATION = HEP-TH 9908142;%%

\bibitem{FrohlichGawedzky}
J.~Frohlich and K.~Gawedzki,
  ``Conformal field theory and geometry of strings,''
  arXiv:hep-th/9310187.
  %%CITATION = HEP-TH 9310187;%%

\bibitem{FGR}
J.~Frohlich, O.~Grandjean and A.~Recknagel,
  ``Supersymmetric quantum theory, non-commutative geometry, and
  gravitation,''
  [arXiv:hep-th/9706132].
  %%CITATION = HEP-TH 9706132;%%


\bibitem{LizziSzaboprl}
F.~Lizzi and R.~J.~Szabo,
  ``Target space duality in noncommutative geometry,''
  Phys.\ Rev.\ Lett.\  {\bf 79}, 3581 (1997)
  [arXiv:hep-th/9706107].
  %%CITATION = HEP-TH 9706107;%%

\bibitem{LizziSzabodual}
F.~Lizzi and R.~J.~Szabo,
  ``Duality symmetries and noncommutative geometry of string spacetimes,''
  Commun.\ Math.\ Phys.\  {\bf 197}, 667 (1998)
  [arXiv:hep-th/9707202].
  %%CITATION = HEP-TH 9707202;%%

\bibitem{LLSstring}
G.~Landi, F.~Lizzi and R.~J.~Szabo,
  ``String geometry and the noncommutative torus,''
  Commun.\ Math.\ Phys.\  {\bf 206}, 603 (1999)
  [arXiv:hep-th/9806099].
  %%CITATION = HEP-TH 9806099;%%

\bibitem{AA-HDD}
  I.~Antoniadis, N.~Arkani-Hamed, S.~Dimopoulos and G.~R.~Dvali,
  ``New dimensions at a millimeter to a Fermi and superstrings at a TeV,''
  Phys.\ Lett.\ B {\bf 436} (1998) 257
  [arXiv:hep-ph/9804398].
  %%CITATION = HEP-PH 9804398;%%

\end{thebibliography}
\end{document}